 \documentclass[10pt, a4paper]{article}
 \usepackage{graphicx}                    
 \usepackage{color}                       
 \usepackage{url}                         
 
\begin{document}



Title: Latitudinal structure and dynamic of the photospheric magnetic field

Authors: E. A. Gavryuseva (Institute for Nuclear Research RAS)

Comments: 22 pages, 10 Postscript figures

 \begin{abstract}
    Analysis of the structure and dynamics of the
    magnetic field of the Sun is  fundamental  for
    understanding of the origin of solar activity and
    variability as well as for the study of solar-terrestrial
    relations.  
    Observations of the large scale magnetic field
    in the photosphere taken at the Wilcox  Solar Observatory
    from 1976 up to 2007
    have been analyzed to deduce  its
    latitudinal and longitudinal structures,
    its differential rotation,
    and their variability in time. 
    This paper is dedicated to the analysis  and dynamics of the latitudinal structure 
    of the solar magnetic field over three solar cyles 21, 22, 23.
   The main results discussed in this paper are the following:
   the large scale latitudinal structure is  antisymmetric and composed of
   four zones with boundaries located at the equator, -25 and + 25 degrees,
   stable over 10-11 years with a time delay
   of about 5-6 years in near-equatorial zones.
   The variability and North-South asymmetry of polarity waves
   running from the equator to the poles with 2-3 - year period
   was studied in detail.
  \end{abstract}
 \vspace{1pc}
 \noindent\textit{Keywords\/}: Sun; solar variability; magnetic field; latitudinal structure; solar cycle.
        \section{Introduction}
      General panoramas of the solar-terrestrial relationship
      can be revealed only by comparison of global characteristics 
      of the processes on the Sun with a broad set of parameters of
      the solar wind and geophysical indexes measured during a long interval of time.
      For this aim the solar magnetic field is
      one of the key characteristics involved in all the main
      processes of solar dynamics.

    Different aspects of the magnetic field of the Sun has been studied 
    (see, for example, the reviews of
    Bumba, 1976;
    Rabin et al., 1991;
    Stix, 2004
    and references there).

 Since the beginning of sunspot magnetic field
 spectroscopic observations the following characteristics,   
 known as Hale's law,
 have been deduced:
  \begin{enumerate}
  \item
    Each sunspot group (and its associated active region)
    is permeated by a bipolar magnetic field.
    The leading (according to solar rotation) and
    the following opposite polarity regions
    are generally of the same sign during each 11-year activity cycle.
  \item
    These polarities are opposite in the northern and
    southern hemispheres.
  \item
    The  polarities change from one activity cycle to the next,
    resulting in a solar magnetic cycle of about 22 years:
    during odd (even) numbered 11-year cycles the leading polarity
    is positive (negative) in the northern hemisphere.
  \end{enumerate}

    According to magnetic field observations made since
   the introduction of the Babcock magnetograph
   (Babcock, 1953;  Babcock and Babcock, 1955;
    Babcock, 1959; Babcock, 1961)
   the polarities of the sub-polar zones are opposite;
   they change around the maxima of the activity cycles;
   before the reversion during odd (even) cycles the polarity
   in the northern sub-polar regions is positive (negative).

   The study of topology and dynamics of the 
   solar magnetic field  is necessary
   because its large scale structures strongly influence
   the coronal field, solar wind propagation and geomagnetic perturbations.


  This paper is focused on the study of the latitudinal structure
  of the large scale photospheric magnetic field (SMF)
  and their variability through solar activity cycles.

   The data used are described in chapter 2.
   Chapter 3 is dedicated to the analysis of the latitudinal  structure,
   chapter 4 to the North-South asimmetry
   of the  photospheric magnetic field.

  The next papers by Gavryuseva E. (2006d, 2006e, 2006f, 2018a, 2018b, 2018c, 2018d)
  are  devoted to the study of the rotation rate SMF (Gavryuseva, 2006d, 2018a), 
  the longitudinal structure (Gavryuseva, 2006e, 2008b, 2018b) and connections of the
  present findings  with
  solar wind and  geomagnetic characteristics (Gavryuseva, 2006f, 2018c, 2018d).
  
  \section{Data}
  The  Wilcox Solar Observatory  (WSO) data
  from http://wso.stanford.edu/ have been selected
  as  one of the best and longest data
  sets of the large scale photospheric  solar magnetic field (SMF)
  observations covering  the solar activity cycles
  No 21, No 22, and No 23
  (Scherrer et al., 1977).

  The line-of-sight component of the photospheric magnetic field is
  measured  by  the WSO's Babcock solar magnetograph
  using the Zeeman splitting
  of the 525.02 nm Fe I spectral line. Data is available beginning
  May 27, 1976 corresponding
  to the beginning  of Carrington Rotation (CR) 1642.
  The data set analyzed covers the interval up to CR 2061.
  Daily magnetograms are taken with a three  arc minute
  spatial resolution which corresponds to
  about 10 heliographic degrees at the center of the solar disc.
  The rotational grid of the available data is made of 30
  equal steps  in  sine of latitude $\theta$ (sin($\theta$))
  where $\theta$ is changing from   75.2 North to 75.2 South degrees
  at the latitudes $\theta$ of $\pm75.2$, $\pm64.2$, $\pm56.4$, $\pm50.1$,
  $\pm44.4$,  $\pm39.3$, $\pm34.5$,  $\pm30.0$,
  $\pm25.7$,  $\pm21.5$, $\pm17.5$,  $\pm13.5$,  $\pm9.6$,  $\pm5.7$,
  $\pm1.9$ degrees, and of 5 degree steps in heliographic longitude.
  Obviously the spatial resolution is declining with increasing latitude.
  The regions beyond 75 degrees of latitude are not resolved at all due to
  the large size of the magnetograph   aperture.
  Each longitudinal value is a weighted average of all the observations made
  in the longitudinal zone within 55 degrees around central meridian.
  The noise level of each measurement is less than 10 micro Tesla 
  (Hoeksema, 1984).
  
 \section{Latitudinal Structure}
  The magnetic field involved in the active regions has a very high intensity.
  Because of this the SMF intensity (SMFI)
  is an index closely related to the solar activity.
  In  Fig. 1 the intensity of the photospheric field  is plotted
  as a function of time $t$ and latitude $\theta$
  averaged over 1 CR (upper plot) and over 1 year (bottom plot).
  Yellow   colors indicate high values of SMFI.
  These plots look very similar to the famous 
  "Butterfly Diagram" of the sunspot distribution over heliographic latitudes
  through the last three solar cycles.
  The polarity of the SMF is not taken into account in this case.
  Any averaging does not change the general  shape 
  of the SMFI($\theta$, $t$) distribution
  while it makes difference  for a magnetic field mean
  which considers the sign of the polarity.

  To study the SMF structure it is necessary
  to take into account the intensity and
  the polarity of the field.
  Let us call the SMF mean over one or
  several full rotations of the Sun as
 the mean latitudinal of the field.
  Stability of the latitudinal field polarity
  and its value over several rotations 
  and at the neighboring latitudes should be considered
  as evidence of the  non random character of the latitudinal field.
  If such field exhibits some clear structure
  then it is important for the understanding
  of the origin of the solar activity.
  
  \subsection{Zonal Structure Stable over Solar Cycles}
    The search for long term latitudinal magnetic 
    field structures should be performed
    by averaging the SMF($\theta$, $t$) around the Sun
    at each latitude $\theta$
    over one or more rotations.
    In the upper  plot of Fig. 2  the mean over 1 Carrington  rotation
    of the solar magnetic field is plotted as a function of time and latitude $\theta$.
    In the bottom plot the yearly running mean SMF with 1 CR step is plotted.
    Yellow and red  (blue) colors indicate  positive (negative) polarity.
    In the upper plot the 0-level of magnitude is indicated by a black line,
    in the bottom plot there are additional levels corresponding to 
    $\pm 50$  micro Tesla  of the yearly mean SMF.
    This plot is an immediate result  of the temporal averaging of the SMF, and it
    agrees with the SMF temporal behavior  using different observational data
  (Bumba, 1976;
   Hoeksema, 1984;
   Howard et al., 1991), etc.
     The origin of the solar cycle and the relationship between
     the  large scale solar magnetic field, surface flows and
     local activities have been widely  discussed
    (Parker, E.N., 1955a, 1955b, 1975, 1988;
     Bumba and Howard, 1965, 1969;
     Howard, 1974a, 1974b, 1996;
     Duvall,         1979;
     Stix,           1981;
     Howard and LaBonte, 1980, 1981;
     DeVore et al., 1984;
     McIntosh and Wilson, 1985;
     Sheeley et al., 1985;
     Sheeley et al., 1987;
     Bogard,         1987;
     Snodgrass, 1986, 1987a,b;
     Rabin   et al., 1991;
     Murray and Wilson, 1993;
     Benevolenskaja, 1996;
     Erofeev,        1996, 1999;
     Obridko and Shelting, 1999;
     Tikhomolov E., 2001;
     Tikhomolov E., Mordvinov V., 2001;
     Ossendrijver, 2003;
     Stix, 2004,  etc.).

    The plots of Fig. 2 clearly show the existence of
    a global $4-$zonal latitudinal structure
    stable over about 11 years having the following properties.
     \begin{enumerate}
     \item
     Four zones can be singled out:
     two pre-equatorial and two sub-polar with latitudinal boundaries
     $\theta$ equal to about  $ +25, 0$ and $-25$ degrees.
     \item
    This large scale zonal structure is antisymmetric relative to the equator
    (the polarities are  opposite in the two solar hemispheres).
     \item
     The polarity is stable over about  11 years.
    The sub-polar zones change their polarity around the maximum
    of  solar activity cycle
    often with some shift in time between the two hemispheres
    as seen in Fig. 2.
    The annual variations of the sub-polar magnetic field
    seen in the upper plot of Fig. 2 are  due to the Earth's orbital motion.
     \item
     While the boundary of the pre-equatorial zones are relatively stable over 10-11 years
     the change of the polarity in sub-polar zones starts first at $\pm 25$ degrees,
     later on it is changing at higher latitudes and in 2-3 years (during solar activity maximum)
     the reversion takes place in poles.
     \item
    The pre-equatorial zones have the same polarity as the leading part
    of  most of the activity regions there; their polarities change
    from one cycle to the next  5-6 years
    before the reversion in the sub-polar zones.
     \item
    The latitudinal structure is reconstructed
    with a periodicity of 20-22 years.
     \end{enumerate}
    Let us call this SMF topology as $4-$zonal 22-year periodical
    latitudinal structure or, shortly, 4-zonal topology.
    This  4-zonal topology   is qualitatively described by the spherical function
    $Y^m_l(\theta, \phi)$, with spherical degree  $l=3$
    ($l$  is the total number of nodal lines)
     and azimuthal order $m=0$
    ($m$ is the number of nodal lines around the equator),
     modulated by the 22-year periodicity
     (Gavryuseva and Kroussanova, 2003).
    A similar model was  suggested by
     Benevolenskaja  (1996, 1998),  see also
    (Stenflo, 1974;
     Stenflo and Vogel, 1986;
     Stenflo and Gugel, 1988;
     Erofeev,           2001;
     Lawrence et al.,   2004;
     Cadavid et al.,    2005).

     The boundaries of the pre-equatorial zones are unexpectedly stable.
     They are very slowly approaching to the equator, significantly slower
     than the drift of the SMFI maximum
     which shows the "Butterfly" migration 
     visible in Fig. 1.
     This is illustrated better by Fig. 3
     where   the zero lines of the polarity inversion
     are plotted  by   continuous lines for the comparison with
     the latitudes weighted with the SMFI between 0 and
     $\pm 50$ degrees  shown by points without averaging
     (upper plot) and by dotted lines for  the latitudes
     weighted with 1 CR mean of the SMFI (bottom plot) in each hemisphere.
     Additionally  in the bottom plot
     the latitudinal positions of the maxima of the SMFI mean over 1 CR
     for the each hemisphere  are shown.

     In the following plots 4, 6, 7, 8 and 10 the latitudes
     weighted with 1 CR mean of the SMFI are plotted by dotted lines.

    While the polarity of the four zonal structure
    is quasi stable over 11 years,
    the intensity of the SMF is changing significantly with
    the 11-year period.
    The amplitude of the 1 year mean solar line-of-sight
    magnetic field  variability  (MSMF)
    is 135-170 micro Tesla   in the sub-polar zones
    at the latitudes of $\pm 75$ degrees.
    The amplitude in the near-equatorial zones is maximal
    at 10-15 degrees latitudes and it is
    equal to 150-200 micro Tesla.

    A quasi 20-year periodicity of the auto-correlation
    of the MSMF distribution for different latitudes
    as a function of time shift in years
    is illustrated by Fig. 4.
    Orange (blue) colors correspond to the positive
    (negative) coefficients of the MSMF($\theta$, $t$) auto-correlation.
    An important characteristic of this auto-correlation
    is its very low level at the
    active latitudes, around zonal boundaries
    where the amplitude of the variability
    of the SMF intensity is the highest.

   The origin of four zonal structure has been studied by Gavryuseva (2008a),
   and it was concluded that the magnetic field of middle intensity (from 5 to 2000 micro Tesla)
   is responsible for this structure due to the North-South asimmetry of the positive and negative
   components of the magnetic field measured in the photosphere.

    For geophysical applications the 
    North-South asymmetry  of the solar activity  is important. This problem 
    can studied by the direct comparison
    of the magnetic field at the latitudes $\theta$ and $-\theta$ in
    both the hemispheres.
    In the upper plot of Fig. 5 the solar magnetic field mean 
    over 1 CR  is shown again as a function of time and latitude.
    Orange (blue) colors indicate  positive (negative) polarities.
    The contours in Fig. 5 correspond to zero level and to $\pm50$ micro Tesla .

     The antisymmetrical part  is deduced for each $\theta$
     as a difference between the SMF rotational means
     in the opposite hemispheres at each latitude:
    $$ 0.5* (SMF(\theta, t) - SMF(-\theta, t)).$$
    These differences are shown in the middle plot.
     This  antisymmetrical to the equator SMF part
     shows the mean level of the magnetic field of opposite polarity
     at  each latitude in the two hemispheres.
     It clearly shows
     the common characteristics of the temporal behavior
     of the magnetic field of opposite polarity in the northern
     and in the southern hemispheres and demonstrates the structure
     of the antisymmetrical field.
     The middle plot clearly confirms the presence of the 4-zonal structure.
     Four zones are visible even better on the middle plot
     than on the upper plot,
     because the antisymmetric parts which have the same amplitude
     in both hemispheres give the main  contribution to this structure.

    The symmetrical part is deduced summarizing the rotational means
    of the solar magnetic field in both hemispheres
    at the same latitudes:
    $$ 0.5*(SMF(\theta, t) + SMF(-\theta, t)).$$
    By definition this part is symmetrical to the
    equator:
    high (low) values correspond to the presence of prevailing
    (non prevailing)   polarities.
    The  North-South symmetrical part of the SMF rotational means 
    is shown in the bottom plot.

     The bottom plot shows an interesting feature of the SMF topology:
     the presence of polarity streams moving from low to high
     latitudes. This high frequency component was noted in
     different solar characteristics, for example, such as neutrino flux,
     radius, $p-$mode frequencies
      (Gavryuseva and Gavryusev, 1994, 2000;
       Delache et al.,    1993;
       Gavryusev et al.,  1994;
       Kane, 2005)
     and a mean-field  axisymmetrical bi-components
     dynamo model was suggested to describe it
      (Benevolenskaja, 1996, 1998).
    The multi modal approach was used by
       Stenflo  (1974),
       Stenflo and Vogel  (1986),
       Stenflo and Gugel  (1988),
       Lawrence et al.    (2004),
     and the relationship between solar activity and the
     short-term variability of the axisymmetric SMF was
     investigated by
       Erofeev  (2001).

     Such an interesting phenomenon should
     be investigated in all the details through several cycles
     and not only the symmetrical part of these streams.
     To study the phase relationship between them in
     both hemispheres it is necessary to use the residuals of the SMF.

 \subsection{Zonal Structure Running Through Latitudes}
   The four zonal quasi-stable structure include
   sub-structures of  lower amplitude if they exist.
   To search them it is necessary to apply a filter 
   to suppress the 4-zonal structure as much as possible.
   Doing this 
   an additional latitudinal topology of the SMF with the polarity
   stable over about one year
   and running through the latitudes to the poles 
   can be  easily revealed. First Howard and LaBonte  (1981)
    described discrete poleward streams.  Then
      Stenflo  (1994),
     Benevolenskaja    (1996, 1998);
     Erofeev    (2001),    Lawrence et al.  (2004) studied 
   the axisymmetric variability on the time scale of about 2 years.
   This structure could be called a  polarity wave
   running through latitudes with quasi 2-year periodicity 
   or shortly zonal wave topology.

   This dynamical structure is clearly visible in Fig. 6
   where the deviation of the yearly mean of the SMF
   from the 2-year SMF mean  at each latitude  with 1 CR step is plotted.
   In other words this deviation
   is a filtered magnetic field (FMF). 
   The FMF looks similar to
   the symmetrical part of the SMF means
   (to the North-South SMF means) but it is not necessarily symmetric,
   and a phase relationship between the   FMF($\theta$, $t$) and
   FMF(-$\theta$, $t$) could be studied
    (Gavryuseva, 2005, 2006, 2006a,b,c;   Gavryuseva \& Godoli, 2006).
     \begin{enumerate}
     \item
     The main difference from the 4-zonal structure
     stable over 10-11 years
     discussed in the previous subsection
     is the running wave character of the polarity which is the same
     for all longitudes over one year and 
     which is moving through latitudes.
     \item
     The amplitude of the running waves is of the order
     of 40  micro Tesla  but it reaches
     90  micro Tesla  at the latitudes of 20-25  degrees South. 
     This value is  half of that in the 4-zonal topology.
     \item
     These polarity running waves have a period of about 2-3 years, and
     they need 2.5-3 years to run from the equator to the poles
     with a velocity of about 40 km/h.
     \end{enumerate}

  The presence of such polarity waves can
  explain the "double"  maximum of the
  activity cycle. Actually 11 years include five full 2-year periods,
  but some of the waves are partly masked by the interference
  with the polarity waves turning back from the poles.
  Let us consider for example the wave of negative polarity which appears
  on the equator in 1980 and the following positive polarity wave in 1981.
  They move to the southern and northern poles and reach 75-degrees in 
  1982 and 1983. The estimated time needed to reach the poles is
  about $2.5-3$ years  (above  75 degrees the waves
  could be only interpolated because
  these regions are not resolved).

   It is possible to note in Fig. 6 an interesting phenomenon:
   the presence in 1985 on the middle latitudes in the northern hemisphere 
   of a clearly visible positive polarity  and of the following negative one.
   They smoothly move to the South,
   cross the equator in 1987.5--1988  and in 1989--1990,
   reach the southern pole in 1990.5--1991  and in 1992,
   and then  continue the motion,
   but to the northern pole,
   which is reached in 1995.5--1996 and in 1997--1998.

   The   interference of such a pattern with the zonal structure
   produces a  modification of the  global 4-zonal topology
   from cycle to cycle. Some features of the weaker 
   zonal wave topology
   can be recognized  even without the filtering of the
   4-zonal topology whose amplitude is at least twice  stronger.
   Indeed in both plots of  Fig. 2 streams of positive
   and negative polarity   moving from the equator
   to higher latitudes can be clearly seen  (for the 1 CR and 1 year
   mean solar magnetic field in the upper and bottom plots).
   The running wave topology could explain the oscillating character of the polarity
   inversion in the sub-polar zones.

   The periodicity of the drifting latitudinal structure can be studied
   by the auto-correlation
   of the FMF data sets as a function of  time shift
   for each latitude.
   The result of the calculation of the correlation coefficient
   is plotted in Fig. 7.
   The 2-year periodicity is clearly visible.
   The mean correlations over all latitudes in each hemisphere are shown
   at the $\pm 80$ degrees levels (multiplied by factor 30).
   There are maxima at 2, 3, 4, 6, 7.7, 10, etc. years shifts
   in the southern hemisphere;
   and at 3, 5, 7, 8.5, 10.5, etc. years shifts in the northern hemisphere,
   while the high correlation FMF and phase reconstruction
   over whole Sun takes place  20-21 years later.
   The FMF zonal wave topology 
   has a quasi 20-year periodicity.
   
  \section{North-South Correlation of Solar Magnetic Field Variability}
  It is interesting to study the phase relation
  between the variability of the
  solar magnetic field in the northern and
  southern hemispheres  at the same latitudes.
  The   coefficient of correlation
  between northern and southern MSMF as a function of time delay
  for different latitudes  is  shown in Fig. 8.
  Orange (blue) colors correspond to the positive
  (negative) coefficient of the MSMF($\theta$, $t$)
  and MSMF($-\theta$, $t$) correlation.
  Strong anti-correlation at zero delay (less than -0.9)
  is presented over almost  all latitudes in 4 zones
  (except around  the zonal boundaries).
  A quasi 22-year periodicity is presented
  in the North-South MSMF correlation.
  The correlation is the lowest between the MSMF on the zone boundaries
  of the  4-zonal structure.

  Strong 22-year periodicity of the North-South SMF correlation
  covers all shorter and weaker  variations which we discussed
  in the previous section.
  The filtered magnetic field (FMF) should be used 
  for the analysis of the short term North-South SMF correlation
  at each pair of latitudes: $\theta$ and -$\theta$.

  Fig. 9 is an illustration of the correlation between
  subsets of the   2-year long FMF running through 29 years
  with 1 CR step located at the same latitudes in northern and southern hemispheres:
  FMF($\theta$,  $t_{i}:t_{i+1}$) and FMF($-\theta$, $t_{i}:t_{i+1}$).
    Orange (blue) colors correspond to the positive (negative) correlation
    between the northern and southern FMF variability.
   High correlation around the equator is attributed
   to the fact that SMF observational data on the
   neighboring latitudes are not completely independent at low latitudes.
  But the correlation coefficient higher than 0.9
   on the latitudes $\theta$ above 10 degrees confirms
   that the FMF variability   is synchronized  in both hemispheres in the
   middle and sub-polar zones during some intervals of time
   well visible in Fig. 9 as yellow color regions.

   The periodical character of the North-South synchronization of the short term
   variability of the FMF is demonstrated in Fig. 10 for
   the correlation between the full FMF data sets 
   related to  $\theta$ and - $\theta$
   latitudes as a function of time shift and $\theta$.
    Orange (blue) colors correspond to the positive (negative)
   correlation coefficient.

   The 2-year period in the FMF North-South correlation is
   presented through all latitudes. Dashed line at the level of 80 degrees
   corresponds to the mean correlation coefficient multiplied by 30.
   The highest correlation takes place with the delay of
   about 1.8, 9.4, 11.7, 20 and 22 years.
   The significance of this short term periodicity in the 
   North-South FMF correlation is high because it is calculated
   for the 29-year long data sets.

   Such well synchronized short and long term variability
   of solar magnetic field in both the hemispheres confirms the
   reality of the two basic topologies: four zonal 
   and waves running through latitudes 
   and their periodical character.
   And even more, this synchronization has a quasi-periodical character
   (with about 2 and 10-year periodicities) as it can be deduced
   from Fig. 10.

   This structure of the running waves type is mainly due to the
   magnetic field of middle intensity (from 5 to 2000 micro Tesla
   as it follows from the recent study of   Gavryuseva, (2008a).

   \section{Summary}
  \begin{enumerate}
  \item   
      The latitudinal structure of the solar magnetic field with a
   polarity period of 22 years has been studied
   and compared with the SMF intensity behavior.
   It consists of four
   zones: two high latitude and two near-equatorial zones with
   boundaries around $+25$, $0$ and $-25$ degrees.
  \item
    The presence of polarity waves
    running from the equator to the poles with quasi 2-year period
    during the last three cycles
    has been clearly demonstrated.
 \item
  The North-South asymmetry of the solar magnetic field and
  its short- and long-term variability in time
  have been studied
  and their synchronized character has been shown.
  \end{enumerate}

   The magnetic field of the Sun has highly-organized latitudinal structure over the solar surface
   and over time, 
   
\section*{Acknowledgments}
   I thank very much the WSO team for their great efforts
   in the measurements of the photospheric field.
   Thanks a lot to Prof. L. Paterno and Dr. E. Tikhomolov
   for precise and profitable advises and
   help in preparation of this paper.
   I am very grateful to Prof. B.T. Draine for his help
   in the revision  of this paper.


 \newpage
 \clearpage
      \begin{figure}
 \centerline{
     \includegraphics[angle=90, width=39pc]
         {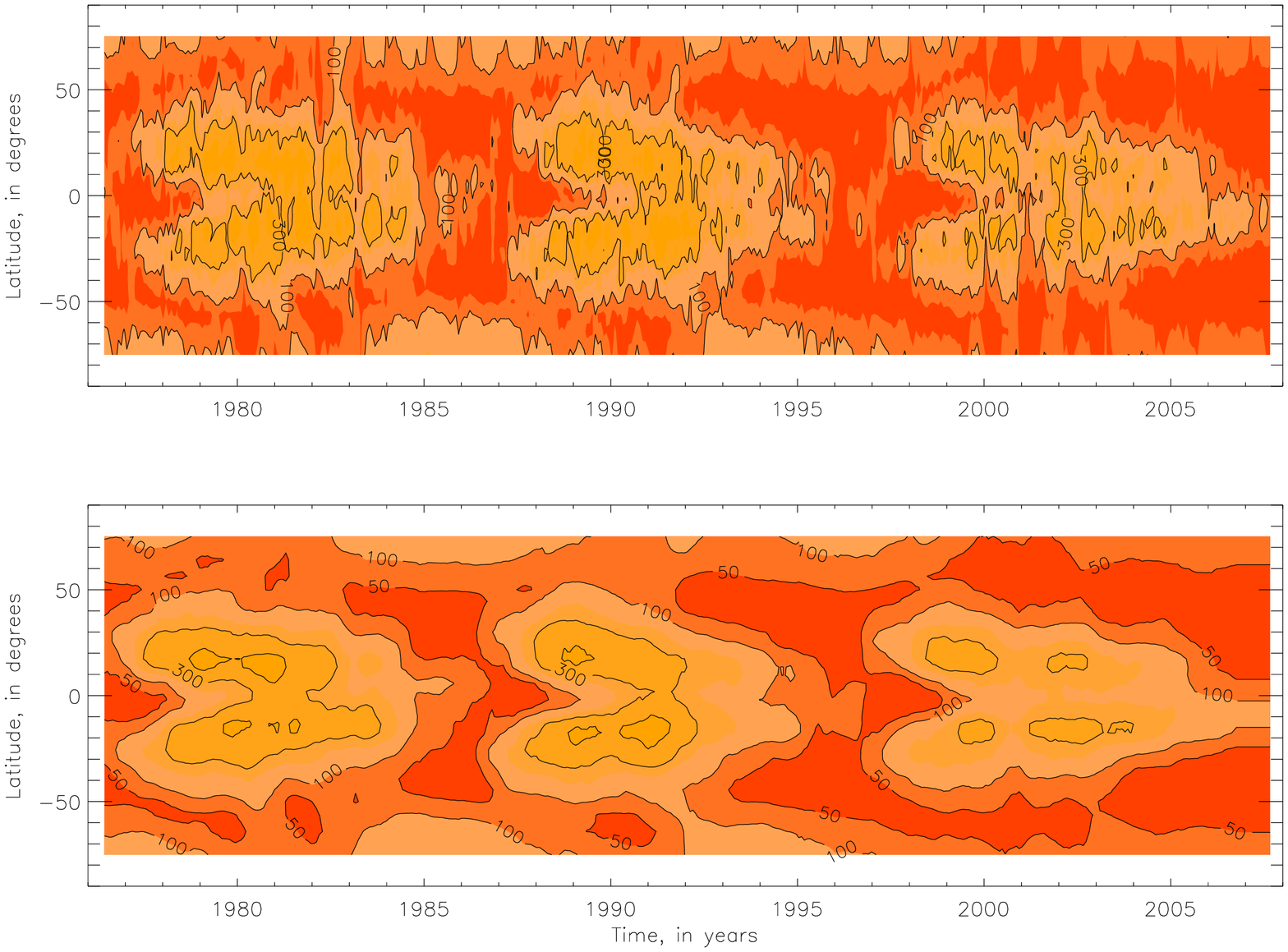}
 }
    \caption
    {
    Distribution of the magnetic field intensity
    in latitude and in time averaged
    over 1 CR (upper plot) and
    over one year  (bottom plot) with 1 CR step.
    Yellow colors indicate high intensity values.
    In the upper plot the contours correspond to the 100 and 300 micro Tesla levels,
    in the bottom plot there is an additional level corresponding to
    $50$  micro Tesla  of the yearly mean SMFI.
    }
   \end{figure}

   \begin{figure}
 \centerline{
    \includegraphics[angle=90, width=39pc]
    {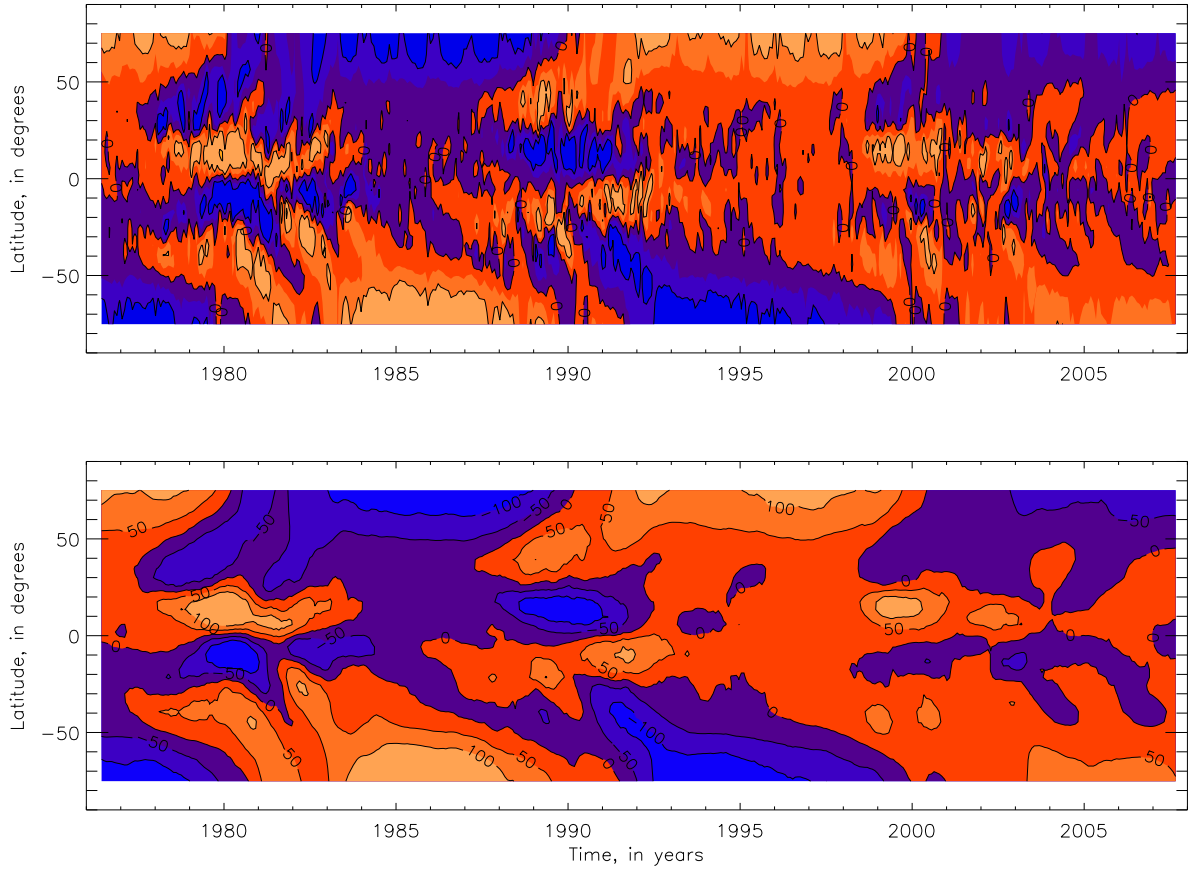}
 }
     \caption
   {
    Distribution of the magnetic field
    in latitude and in time averaged
    over 1 CR (upper plot) and
    over one year  (bottom plot) with 1 CR step.
    Yellow and red (blue) colors indicate positive
    (negative) polarities.
    In the upper plot the contours correspond to the levels
     of 0, $\pm 50, 100, 300$ micro Tesla,
    in the bottom plot there are  additional levels corresponding to
    $\pm 50$  micro Tesla  of the yearly mean SMF.
    In the upper plot 0-level is indicated by a black line.
    In the bottom plot the contours correspond to the levels
     of -50, 0, 50 micro Tesla.
    }
    \end{figure}
  \begin{figure}
 \centerline{
 \includegraphics[angle=90, width=39pc]
    {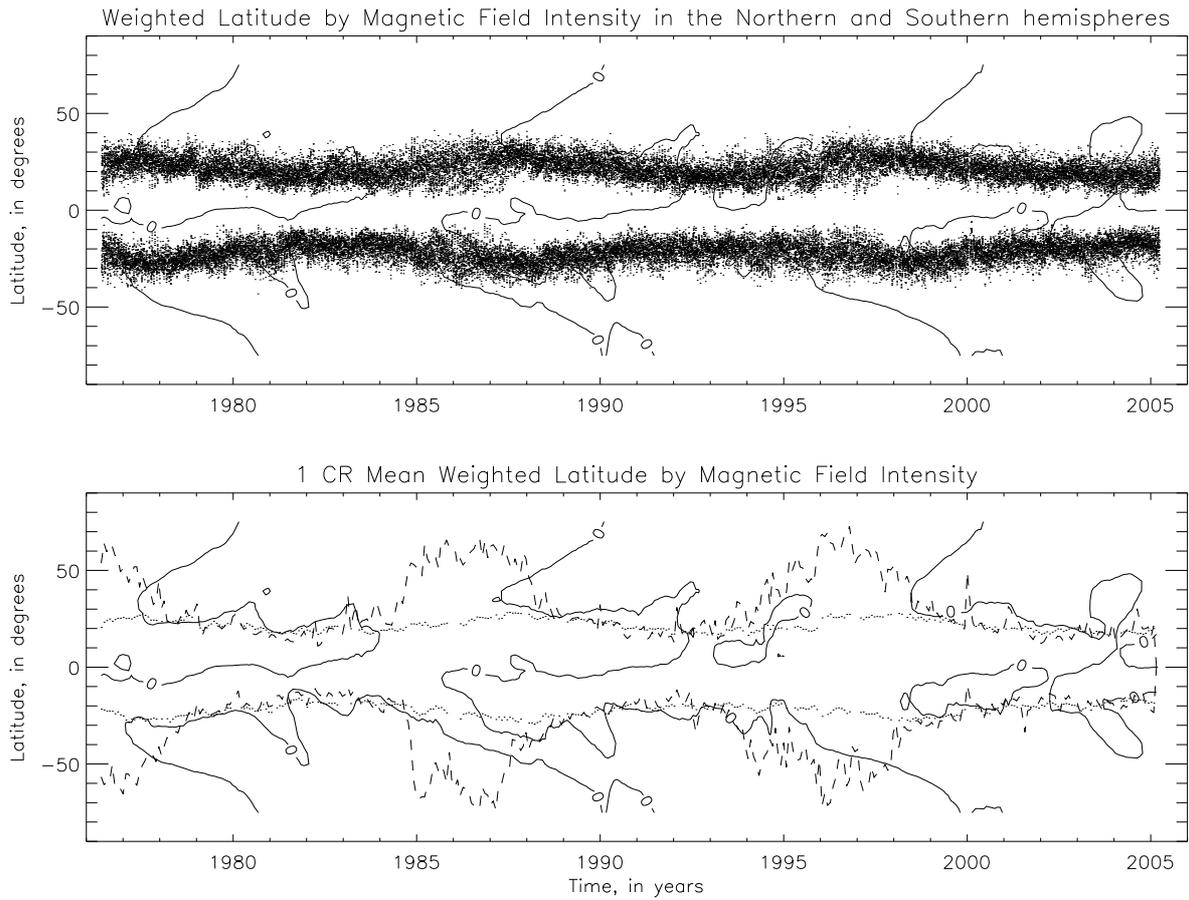}
 }
    \caption
  {
    The polarity inversions are plotted by continuous lines.
    The latitudes weighted with the SMFI between 0 and
    $\pm 50$ latitude degrees are shown by points without averaging
    (upper plot) and by dotted lines for  the latitudes
    weighted with 1 CR mean of the SMFI (bottom plot).
    In the bottom plot the positions of the maxima 
    of the SMFI mean over 1 CR in both hemispheres are shown by dashed line.
   }
   \end{figure}
  \begin{figure}
 \centerline{
 \includegraphics[angle=90, width=39pc]
    {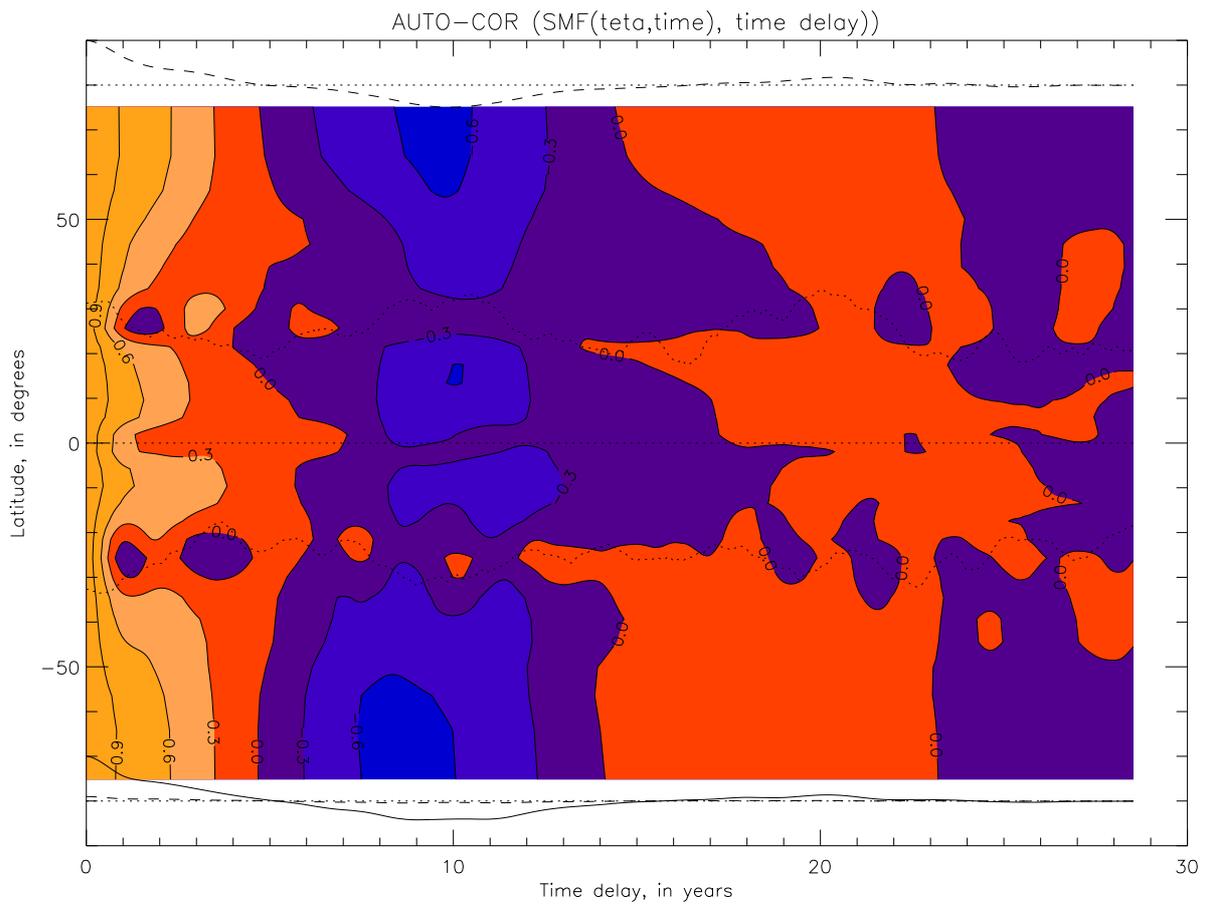}
 }
    \caption
  {
    The auto-correlation of the 1-year mean SMF distribution
    as a function of time for different latitudes.
    Orange (blue) colors correspond to the positive
    (negative) coefficient of the MSMF($\theta$, $t$) auto-correlation.
   }
   \end{figure}
  \begin{figure}
 \centerline{
 \includegraphics[angle=90, width=39pc]
    {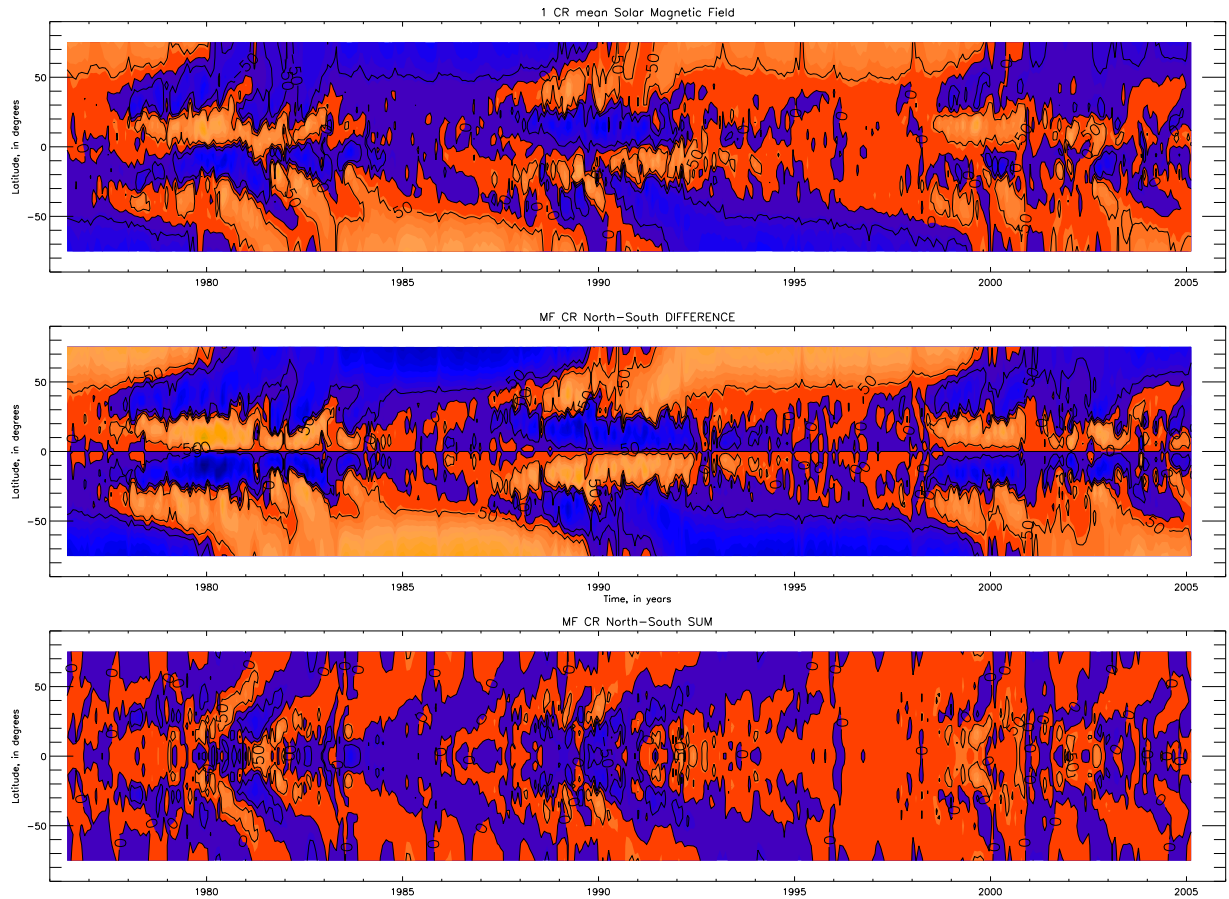}
 }
    \caption
  {
    The mean solar magnetic field
    over 1 CR as a function of time and latitude (upper plot).
    The difference or antisymmetrical part of the SMF 1 CR running means
    is shown in the middle plot.
    The North-South symmetrical part of the 1 CR means
    of the SMF is shown in the bottom plot.
    Yellow  and red (blue) colors indicate
    positive (negative) polarities.
    The contours correspond to zero level and to the $\pm 50$ micro Tesla.
    }
   \end{figure}
  \begin{figure}
 \centerline{
 \includegraphics[angle=90, width=39pc]
    {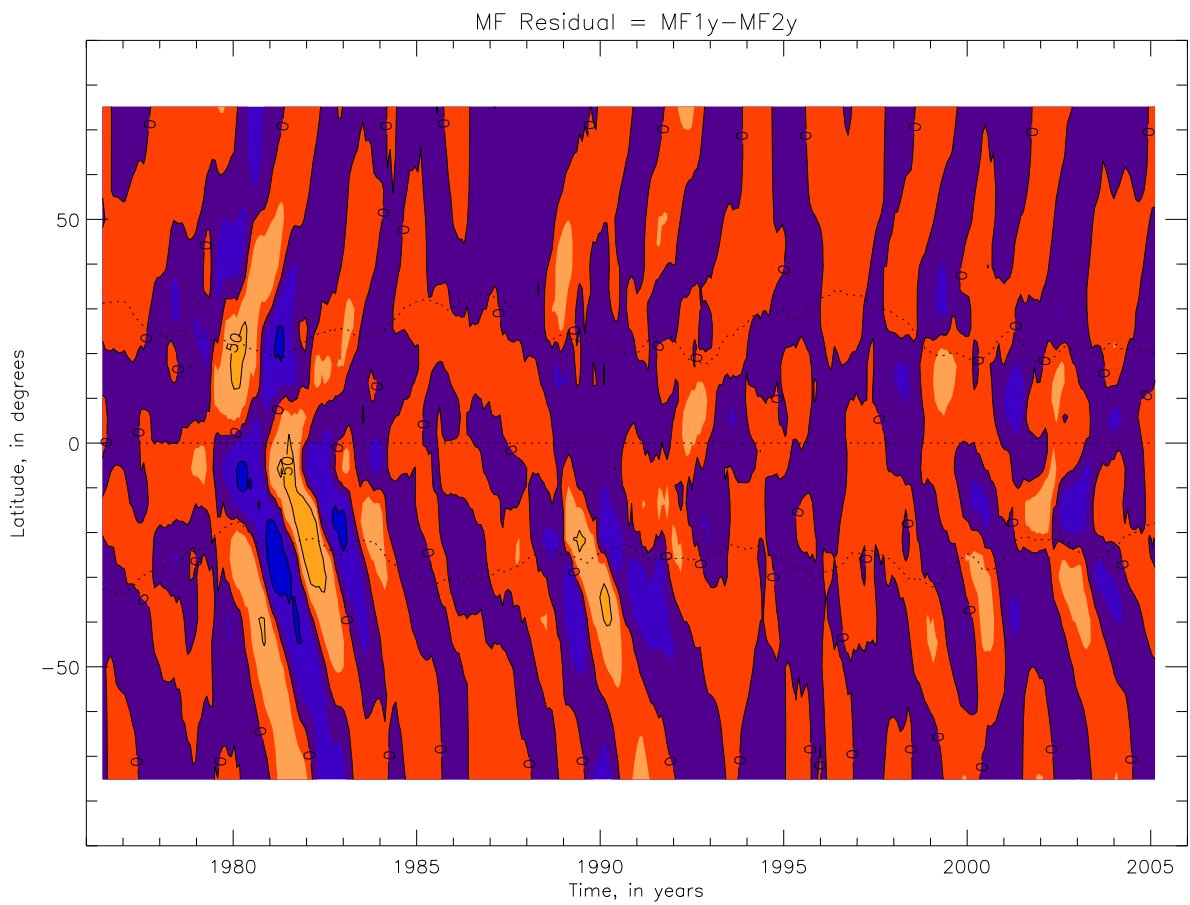}
 }
    \caption
  {
   The deviation of the yearly mean of the SMF from the 2-year mean 
   of the SMF for each latitude  with 1 CR step.
    Orange (blue) colors correspond 
    to the positive (negative) correlation coefficients.
  }
   \end{figure}
  \begin{figure}
 \centerline{
 \includegraphics[angle=90, width=39pc]
    {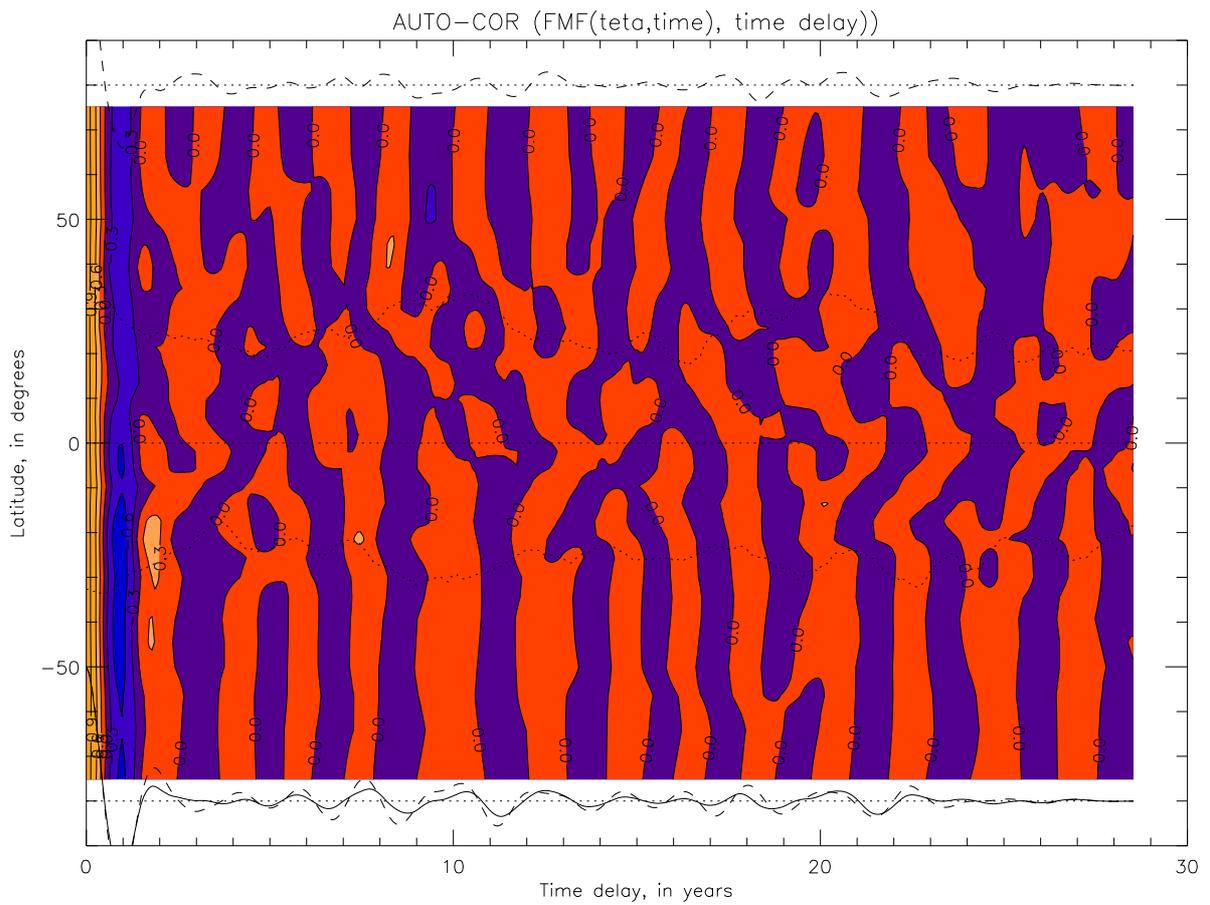}
 }
    \caption
  {
   The coefficient of the auto-correlation
   of the filtered magnetic field 
   data sets as a function of time shift for each latitude.
   Orange (blue) colors correspond 
    to the positive (negative) correlation coefficients.
  }
   \end{figure}
  \begin{figure}
 \centerline{
 \includegraphics[angle=90, width=39pc]
    {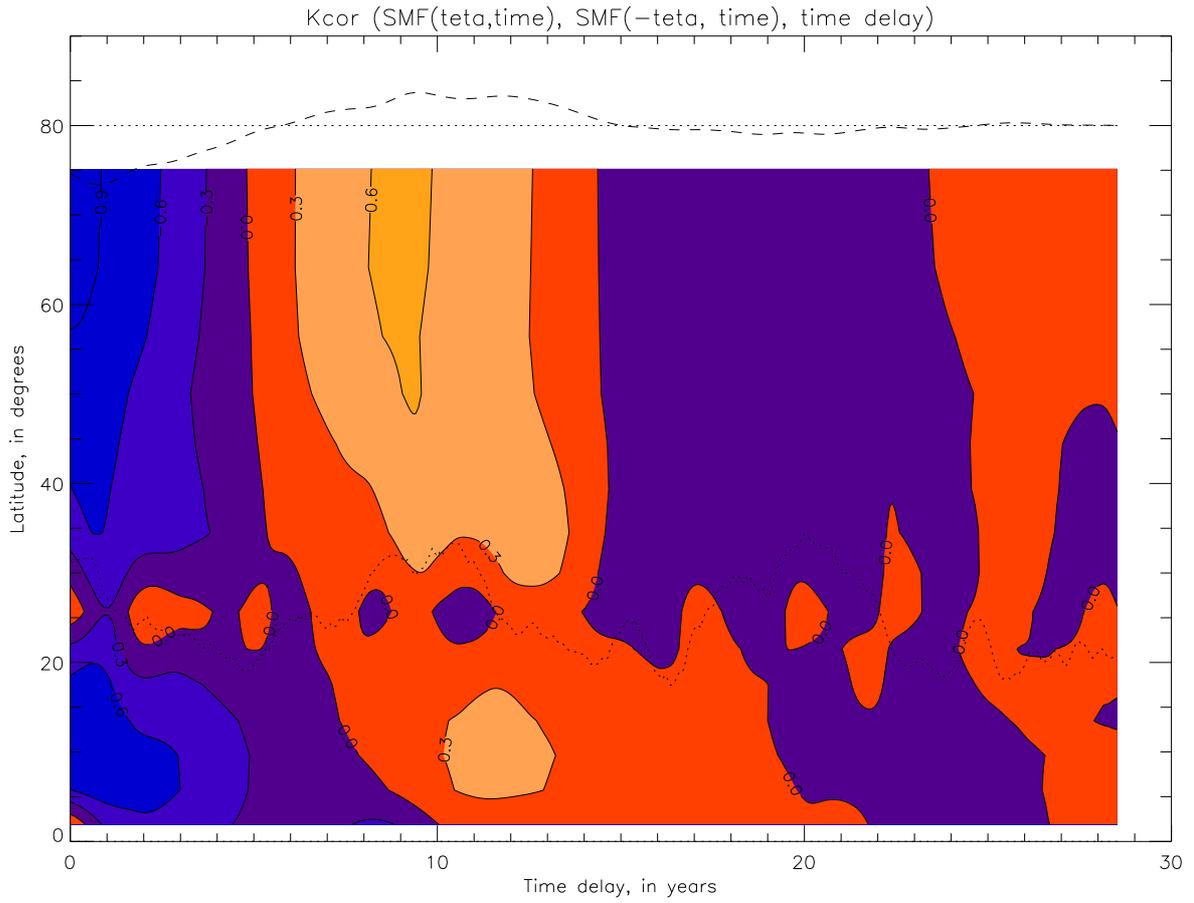}
 }
    \caption
  {
   The   coefficient of correlation of the
   full data sets at $\theta$ and - $\theta$
   latitudes as a function of time shift and $\theta$.
   Orange (blue) colors correspond to the positive 
   (negative) correlation coefficients.
   Dashed line at the level of 80 degrees
   corresponds to the mean correlation coefficient multiplied by 30.
  }
   \end{figure}
  \begin{figure}
 \centerline{
 \includegraphics[width=39pc]
    {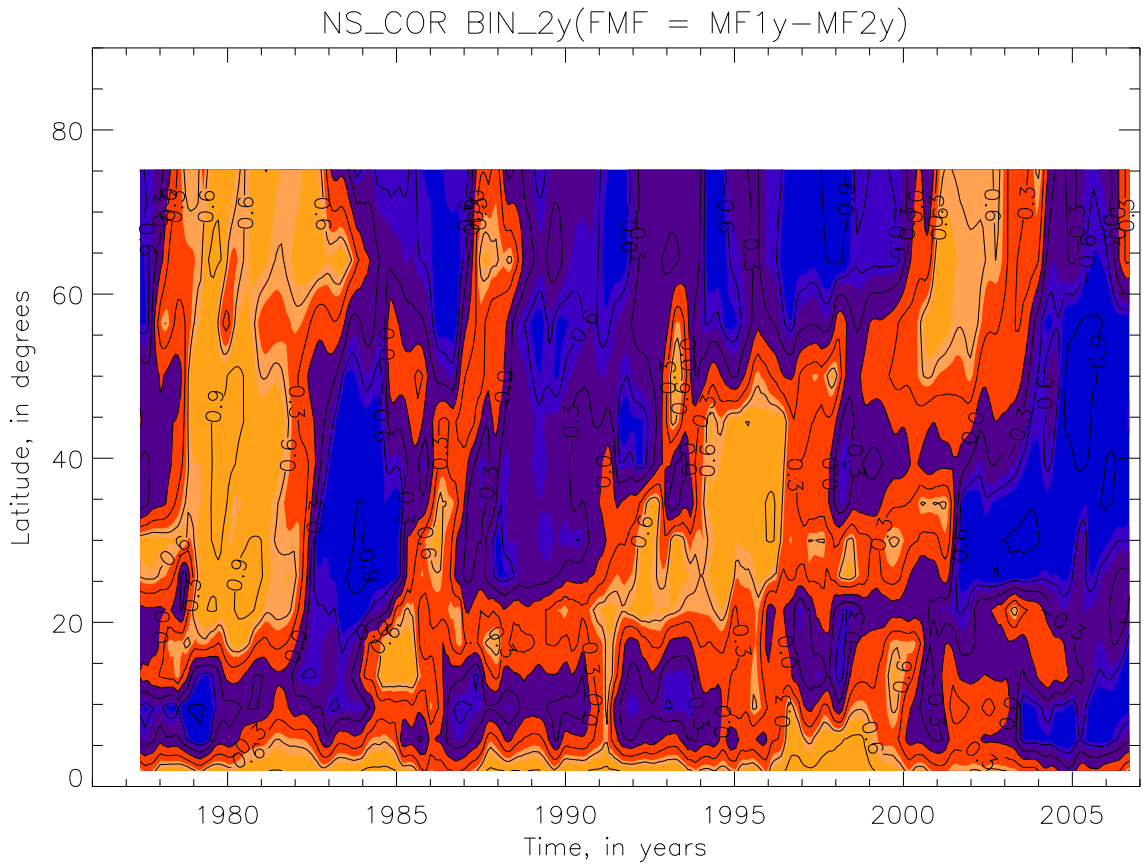}
 }
 \caption
  {
  The   coefficient of correlation of the
  subsets of the FMF  2-year long running through 29 years
  with 1 CR step located at the same latitudes in northern and southern hemispheres:
  FMF($\theta$,  $t_{i}:t_{i+1}$) and FMF($-\theta$, $t_{i}:t_{i+1}$).
   Yellow and orange (blue) colors correspond 
   to the positive (negative) correlation coefficients.
  }
   \end{figure}
  \begin{figure}
 \centerline{
 \includegraphics[angle=90, width=39pc]
    {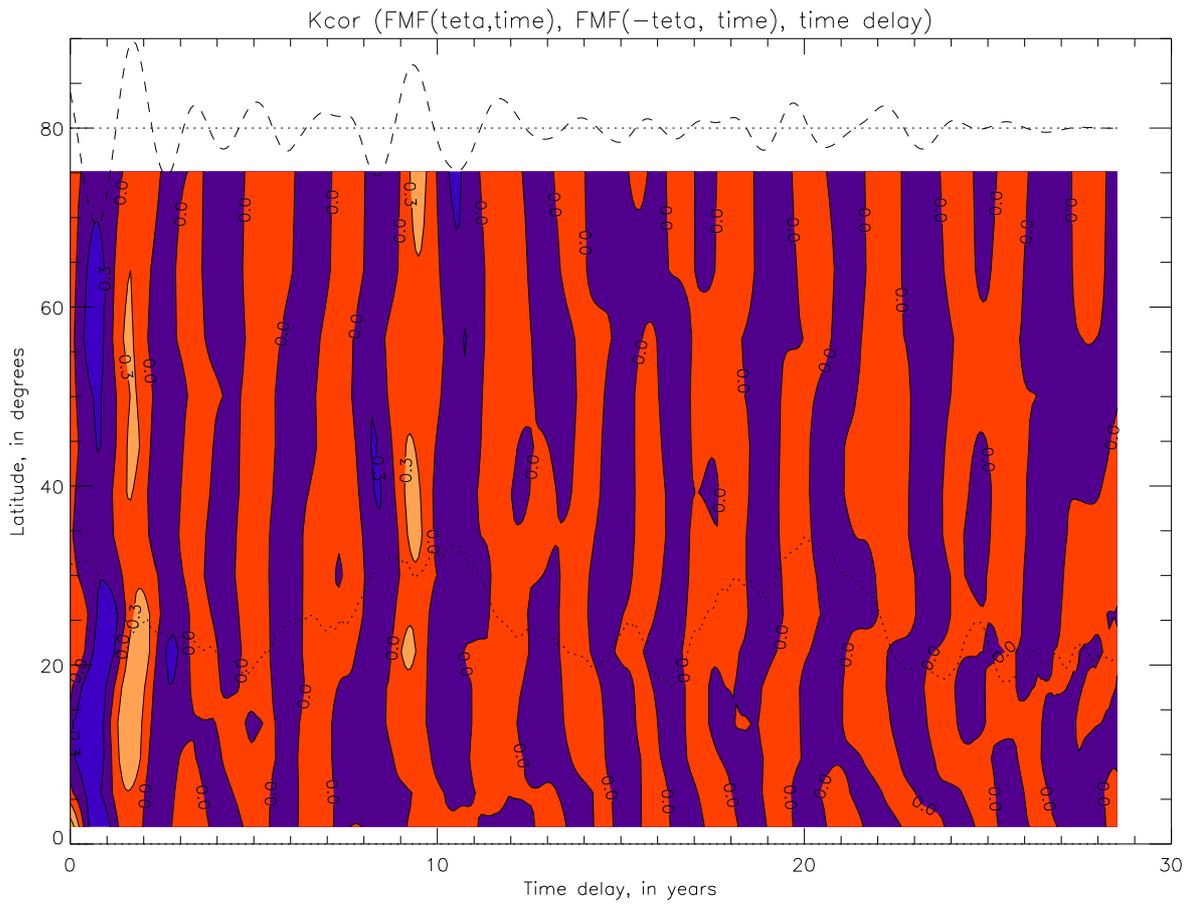}
 }
    \caption
  {
  The   coefficient of correlation
  between the northern and southern FMF as a function of time delay
  for different latitudes.
  Orange (blue) colors correspond to the positive
  (negative) coefficient of the FMF($\theta, t$)
  and FMF($-\theta, t$) correlation.
  }
   \end{figure}

   \end{document}